\documentclass[prd,reprint,superscriptaddress,nofootinbib,longbibliography,aps]{revtex4-1}
\usepackage{graphicx}
\usepackage{longtable}
\usepackage{amsmath}
\usepackage{color}
\usepackage{dsfont}
\usepackage{amssymb}
\usepackage{subfigure}
\usepackage{tabularx}
\usepackage[colorlinks=true]{hyperref}
\usepackage{cleveref}
\usepackage[normalem]{ulem}
\usepackage{slashed}
\usepackage[toc]{appendix}
\usepackage{bm}
\usepackage{braket}
\usepackage{comment}  
\usepackage{orcidlink}

\newcommand{\dd}{\mathrm{d}}

\begin{document}

\title{Cosmological Evidence for Dark Axion--Dark Baryon Interactions from Apparent Phantom Crossing}

\author{Justin Khoury\,\orcidlink{0000-0002-2001-0076}}
\email{jkhoury@upenn.edu}
\affiliation{Center for Particle Cosmology, Department of Physics and Astronomy, University of Pennsylvania, Philadelphia, Pennsylvania 19104, USA}

\author{Meng-Xiang Lin\,\orcidlink{0000-0003-2908-4597}}
\altaffiliation{CITA National Fellow}
\email{mengxiang\_lin@sfu.ca}
\affiliation{Department of Physics, Simon Fraser University, Burnaby, British Columbia, V5A 1S6, Canada}
\affiliation{Canadian Institute for Theoretical Astrophysics (CITA), University of Toronto, 60 St George Street, Toronto, Ontario M5S 3H8, Canada}
\affiliation{Center for Particle Cosmology, Department of Physics and Astronomy, University of Pennsylvania, Philadelphia, Pennsylvania 19104, USA}

\author{Mark Trodden\,\orcidlink{0000-0001-6864-6711}} 
\email{trodden@upenn.edu}
\affiliation{Center for Particle Cosmology, Department of Physics and Astronomy, University of Pennsylvania, Philadelphia, Pennsylvania 19104, USA}

%\date{\today}

\begin{abstract}
    \noindent Interactions between dark matter and dark energy can lead to an apparent phantom-crossing behavior that mimics the expansion history preferred by the latest cosmological observations from DESI baryon acoustic oscillations (BAO), Cosmic Microwave Background (CMB), and Type Ia supernovae (SNe Ia) data. In a previous paper~\cite{Khoury:2025txd}, we proposed a concrete particle physics realization of this idea, consisting of a strongly coupled dark sector in which a dark axion is coupled to dark baryons. In this paper, we investigate this idea further by comparing its predictions to the latest cosmological data. We implement the dark axion--dark baryon interaction model in a Boltzmann code and confront it with CMB, DESI DR2 BAO, and SNe Ia data. For the CMB+DESI DR2+DES-Dovekie combination, the best-fit model improves the fit relative to~$\Lambda$CDM by~$\Delta\chi^2=-14.48$. The preferred solution exhibits a non-monotonic dark-matter mass evolution: the mass decreases between matter-radiation equality and recombination, while increasing over the BAO/SNe-sensitive epoch, leading to an apparent phantom crossing in an effective dark-energy description. Interestingly, the same dynamics produces an Early Dark Energy-like energy injection near matter-radiation equality, but in the data-preferred region this component is too small to raise~$H_0$ enough to substantially reduce the current tension.
\end{abstract}

\maketitle

%%%%%%%%%%%
\section{Introduction}\label{sec:intro}
The~$\Lambda$CDM model provides a successful phenomenological description of a wide range of cosmological observations. Nevertheless, the microscopic nature of dark matter (DM) and dark energy (DE) remains unknown, and several recent observational developments suggest that the dark sector may be richer than in the minimal model.

One long-standing puzzle is the Hubble tension: the value of~$H_0$ inferred from local distance-ladder measurements differs from that inferred from CMB observations assuming~$\Lambda$CDM~\cite{Riess:2021jrx,Planck:2018vyg}. Among the many proposed resolutions, early dark energy (EDE) models are particularly well studied~\cite{Poulin:2018cxd,Lin:2019qug,Smith:2019ihp,Kamionkowski:2022pkx, Poulin:2023lkg}. They add a transient contribution to the total energy density near matter-radiation equality, reducing the sound horizon before rapidly diluting away at later times.

A second, more recent hint comes from DESI BAO measurements. When DESI DR2 BAO data are combined with CMB and SNe Ia observations, the data show a preference for dynamical DE~\cite{DESI:2025zgx,DESI:2025fii}. In the commonly used Chevallier--Polarski--Linder (CPL) parameterization for the DE equation of state,~$w(a)=w_0+w_a(1-a)$~\cite{Chevallier:2000qy,Linder:2002et}, the preferred region has~$w_0>-1$ and $w_a<0$, corresponding to an effective ``phantom crossing'' of~$w=-1$ at redshift $z\sim 0.5$. Extended DESI analyses using more flexible reconstructions, including binned and Gaussian-process descriptions of~$w(z)$, find the same qualitative trend~\cite{DESI:2025fii}. Subsequent SNe Ia recalibrations~\cite{DES:2025sig,Hoyt:2026fve} and multi-probe analyses including DES weak-lensing and galaxy-clustering data~\cite{DES:2026jmi} have tested the robustness of this preference, generally finding a weaker but still intriguing indication for dynamical DE.

A fundamental phantom-crossing behavior~\cite{Hu:2004kh,Feng:2004ad,Guo:2004fq} is theoretically challenging. For a perfect fluid with positive energy density,~$w<-1$ violates the null energy condition. Stable semiclassical realizations of such behavior are difficult to construct: familiar attempts often encounter ghosts, gradient instabilities, superluminality, or the absence of a healthy Poincar\'e-invariant vacuum~\cite{Carroll:2003st,Cline:2003gs,Dubovsky:2005xd,Nicolis:2009qm, Creminelli:2010ba,Creminelli:2012my}. However, the equation of state inferred from cosmological distances need not coincide with the equation of state of a fundamental fluid. In particular, an apparent phantom behavior can arise with a non-minimal coupling between DE and other fields, such as DM~\cite{Melchiorri:2002ux,Huey:2004qv,Das:2005yj,Khoury:2025txd} or gravity~\cite{Carroll:2004hc,Lue:2004za,Dubovsky:2005xd,Ye:2024ywg,Pan:2025psn,Wolf:2025jed,Wolf:2025acj,Garcia-Garcia:2026nzy}. 

In the case of DM-DE interactions, the apparent phantom behavior arises if the data are analyzed assuming separately conserved DM and DE components when, in the true theory, the DM mass evolves through an interaction with a light scalar~\cite{Melchiorri:2002ux,Huey:2004qv,Das:2005yj}. If the DM mass increases with time over the BAO/SNe-sensitive redshift range, an observer who assumes constant-mass DM will attribute the non-standard DM evolution to an effective DE component with~$w_{\rm eff}<-1$.

The challenge is to realize this mechanism in a controlled theory that can simultaneously fit the CMB and low-redshift distance data~\cite{Linder:2025zxb,Caldwell:2025inn}. The CMB strongly constrains the matter density near equality, the sound horizon, the acoustic angular scale, CMB lensing, and the late-time integrated Sachs-Wolfe effect. These constraints make monotonic varying-mass histories difficult to accommodate, and motivate considering a non-monotonic DM mass evolution. 
Some phenomenological treatments of DM-DE interactions have been explored in data analyses after DESI DR2~\cite{Silva:2025hxw,Pan:2025qwy,Li:2025muv,Figueruelo:2026eis,Li:2026xaz,Li:2026asg}.

In Ref.~\cite{Khoury:2025txd}, we proposed a concrete particle physics realization of DM-DE apparent phantom behavior based on dark axion--dark baryon (DADB) interactions in a strongly coupled dark sector. In this model, DM is composed of dark baryons, while the associated dark-QCD axion plays the role of DE. Finite-density corrections to the dark quark condensate induce an axion-dependent contribution to the dark-baryon mass, thereby linking the evolution of DM and DE in a technically motivated way. The scalar field is canonical, so the apparent phantom behavior does not require a fundamental phantom degree of freedom. Related interacting dark-sector realizations of apparent phantom behavior have recently been explored in~\cite{Bedroya:2025fwh,Smith:2025grk,Andriot:2025los,Wang:2025znm,LaPenna:2026avs}.

In this paper we confront the DADB model with cosmological data. We first present a geometric argument showing why the CMB and low-redshift distance data favor a non-monotonic DM mass history: the mass decreases between matter-radiation equality and recombination, but increases over the BAO/SNe-sensitive epoch. We then implement the model, including linear perturbations, in a modified Boltzmann code and compare it with CMB, DESI DR2 BAO, and SNe Ia data. For our default CMB+DESI DR2+DES-Dovekie data combination, the best-fit DADB model improves the fit relative to an independently optimized~$\Lambda$CDM model by~$\Delta\chi^2_{\rm total}=-14.48$. Within the sampled DADB parameter space, the data also prefer a nonzero axion-dark-baryon coupling relative to the decoupled axion-quintessence limit.

Finally, we show that the same dynamics generates an EDE-like component near matter-radiation equality, without introducing an additional field or energy scale. In the data-preferred region this component peaks at~$f_{\rm EDE}^{\rm peak}\simeq 0.008$, which is too small to resolve the Hubble tension, but it illustrates how early- and late-time DE can arise from a single interacting dark sector. The best-fit model also predicts modified growth of structure, making future galaxy clustering, weak-lensing, and redshift-space distortion measurements important tests of the model.

%%%%%%%%%%%

\section{Geometric argument for non-monotonic DM mass evolution}
\label{sec:geo}

It is useful to understand geometrically why an increasing DM mass during the BAO/SNe-sensitive epoch can mimic the expansion history associated with phantom DE, and why the CMB generally requires that the DM mass decrease in the early Universe.
See also the related discussion in~\cite{Weiner:2026sfm}.

We work to first order around a spatially flat fiducial~$\Lambda$CDM cosmology and set~$c = 1$. To isolate changes in the shape of the expansion history, we compare models with the same present-day value of~$H_0$ and the same present-day baryon, DM, and radiation density parameters, with the DE density fixed by spatial flatness. This is simply a normalization choice for the following geometric argument; all cosmological parameters, including~$H_0$, will be varied in the full likelihood analysis. 

Let us define~$\delta\ln H\equiv\ln[H/H^{(\Lambda)}]$, and consider first a constant equation of state $w=-1+\delta w$. At fixed present-day densities, the corresponding first-order perturbation to the Hubble rate is
\begin{equation}
  \delta\ln H^{(w)}(a)=
  \frac{3}{2}\Omega_{\rm DE}^{(\Lambda)}(a)\,
  \delta w\,\ln\frac{1}{a}\,,
  \label{eq:deltaH_w_const}
\end{equation}
where we have set~$a = 1$ at the present time. The superscript $(\Lambda)$ denotes the fiducial $\Lambda$CDM
quantity. More generally, for the CPL parametrization
\begin{equation}
  w(a)=w_0+w_a(1-a)\,,
\end{equation}
the perturbation around $(w_0,w_a)=(-1,0)$ is
\begin{equation}
  \delta\ln H^{(w_0w_a)}(a)=
  \frac{3}{2}\Omega_{\rm DE}^{(\Lambda)}(a)
  \left[
    \delta w_0\ln\frac{1}{a}
    +w_a\left(\ln\frac{1}{a}-(1-a)\right)
  \right],
  \label{eq:deltaH_CPL}
\end{equation}
where $\delta w_0\equiv w_0+1$. Unlike the constant-$w$ case, the CPL perturbation can change sign with redshift, so its effect cannot be inferred from $w_0$ alone.

Uncalibrated SNe constrain the shape of the luminosity distance~$D_L=(1+z)D_M$, while BAO constrain the ratios~$D_H/r_{\rm d}$ and~$D_M/r_{\rm d}$ (or $D_V/r_{\rm d}$ for isotropic measurements), where~$r_{\rm d}$ is the sound horizon at the baryon drag epoch,\footnote{To be precise,~$r_{\rm d} = r_s(z_{\rm d})$, with the sound horizon given as usual by
\begin{equation}
r_s(z)=\int_z^\infty \frac{c_s(z')}{H(z')}\,{\rm d}z'\,,
\end{equation}
where~$c_s= 1/\sqrt{3(1+R)}$ is the sound speed of the photon-baryon fluid, with~$R=\frac{3\rho_b}{4\rho_\gamma}$.} and
\begin{equation}
  D_H(a_{\rm obs})=\frac{1}{H(a_{\rm obs})}\,;
  \qquad
  D_M(a_{\rm obs})=
  \int_{a_{\rm obs}}^1\frac{\dd a}{a^2H(a)}\,.
\end{equation}
For the moment, we hold~$r_{\rm d}$ fixed in order to isolate the low-redshift geometric response. To first order, the transverse comoving distance changes according to
\begin{equation}
  \delta D_M(a_{\rm obs})=
  -\int_{a_{\rm obs}}^1
  \frac{\dd a}{a^2H^{(\Lambda)}(a)}\,
  \delta\ln H(a)\,.
  \label{eq:deltaDM_general}
\end{equation}
Thus a constant phantom perturbation,~$\delta w<0$, lowers the Hubble rate in the past and increases both~$D_H$ and~$D_M$.

Now consider a varying DM mass~$m(a)=m_{\rm N}A(a)$. Normalizing the physical DM density to its present-day value gives
\begin{equation}
  \rho_{\rm DM}(a) =   \frac{\rho_{\rm DM}^0}{a^3}\frac{A(a)}{A_0}\,,
    \label{eq:rho_DM}
\end{equation}
where~$A_0\equiv A(1)$.\footnote{Henceforth, the subscript~``0" indicates present-day values.}
The change in the expansion rate is
\begin{equation}  
  \delta\ln H^{(A)}(a) =   \frac{1}{2}\Omega_c^{(\Lambda)}(a)\,   \delta\ln A(a)\,,
  \label{eq:deltaH_A}
\end{equation}
with~$\delta\ln A(a)\equiv\ln[A(a)/A_0]$. Here we have isolated the direct contribution from the varying DM mass; in the full DADB solution it is accompanied by the scalar-field contribution to the expansion rate. Comparing Eqs.~\eqref{eq:deltaH_w_const} and~\eqref{eq:deltaH_A}, we see that~$\delta\ln A<0$ produces the same geometric response as~$\delta w<0$. Since~$\delta\ln A<0$ means that the DM mass was lower in the past than it is today, an increasing DM mass can mimic phantom expansion over the BAO/SNe-sensitive epoch.

This low-redshift degeneracy cannot, however, be extrapolated arbitrarily far into the past. The CMB tightly constrains both the matter density near equality and the acoustic angular scale
\begin{equation}
  \theta_*=\frac{r_s(z_*)}{D_M(z_*)}\,,
\end{equation}
where~$r_s(z_*)$ is the sound horizon at last scattering.\footnote{In the full numerical analysis,~$z_*$ and~$r_s(z_*)$ are determined self-consistently from the recombination history.} In the~$w_0w_a$CDM model, pre-recombination physics is essentially unchanged, so~$r_s(z_*)$ remains nearly fixed and the CMB acoustic scale anchors the distance to recombination. The effect of the DESI-preferred trajectory, with~$w_0>-1$ and~$w_a<0$, is determined by the full redshift-dependent response in Eq.~\eqref{eq:deltaH_CPL}, rather than by~$w_0$ alone. In the joint DESI and CMB fits, preservation of the acoustic scale leads to a compensating shift toward lower~$H_0$~\cite{DESI:2025zgx,Mirpoorian:2026vxn}.

In our model, by contrast, the pre-recombination sound horizons need not remain fixed. If the late-time increase of the DM mass were extrapolated monotonically back to recombination, the DM density near equality and recombination would be lower than in the corresponding constant-mass cosmology. This would delay matter-radiation equality, reduce the pre-recombination expansion rate, and increase both~$r_s(z_*)$ and~$r_{\rm d}$. These effects are strongly constrained by the CMB. A larger DM mass around equality, relative to the monotonic extrapolation of the late-time branch, compensates for this shift.
Connecting this early enhancement to the subsequent late-time increase requires the DM mass first to decrease and then to turn around, thereby motivating a non-monotonic cosmological history. This expectation will be borne out in our best-fit solution.

%%%%%%%%%%%
\section{The Model}\label{sec:model}
We summarize the dark axion--dark baryon interaction (DADB) model proposed in~\cite{Khoury:2025txd}.
Dark matter is composed of dark baryons, while the associated dark-QCD axion plays the role of dark energy.
Dark axion--dark baryon interactions arise from the finite dark-baryon density correction to the quark condensate~\cite{Cohen:1991nk,Balkin:2020dsr}, which in turn results in a density-dependent contribution~\cite{Hook:2017psm} to the axion potential~\cite{DiVecchia:1980yfw,GrillidiCortona:2015jxo}.

To linear order in density, the effective potential of the dark QCD axion field~$\phi$ reads\footnote{We have added a constant so that the vacuum potential vanishes at~$\phi=0$.}
\begin{align}
\nonumber
V_{\rm eff} (\phi) & =   \Lambda^4 \Bigg\{1 - \left(1 - \frac{2\sigma_{\rm N} n}{\Lambda^4}\right)  \sqrt{1-\xi \sin^2\left(\frac{\phi}{2f}\right)}  \Bigg\} \,,\\
%& + {\cal O}\left( \frac{n^2\sigma_{\rm N}^2}{m_\pi^2f_\pi^2} \right)\, ,
\label{axion pot full}
\end{align}
where~$\Lambda$ is the DE scale ($\sim$~meV),~$f$ is the axion decay constant,~$\xi = \frac{4m_{\rm u}m_{\rm d}}{(m_{\rm u}+m_{\rm d})^2}$ is the dimensionless ratio of quark masses,~$n \sim a^{-3}$ is the DM number density, and~$\sigma_{\rm N}$ is the pion-nucleon sigma term characterizing the dependence of the dark nucleon mass on the average quark mass.
% and $\rho_{\rm DM}$ is .
The coefficient of the square root term flips sign when the DM number density reaches the critical value
\begin{equation}
n_{\rm c} = \frac{\Lambda^4}{2\sigma_{\rm N}}\,.
\label{nc}
\end{equation}
Above critical density~($n > n_{\rm c}$), the potential is minimized at~$\phi = \pi f$; instead, at low density~($n < n_{\rm c}$), it is minimized at~$\phi = 0$. 

From the point of view of DM, the density-dependent term in~$V_{\rm eff}$ is interpreted as an axion-dependent contribution to the dark baryon particle mass:
\begin{equation}
m(\phi) = m_{\rm N} A(\phi)\,,
\label{Adef}
\end{equation}
where
\begin{equation}
A(\phi) \simeq 1 + 2\frac{\sigma_{\rm N}}{m_{\rm N}} \sqrt{1- \xi\sin^2\left(\frac{\phi}{2f}\right)}\,.
\label{DM mass axion}
\end{equation}
Thus~$\sigma_{\rm N}/m_{\rm N}$ is a dimensionless axion-baryon coupling parameter. While the DM number density still redshifts as~$1/a^3$, the physical DM mass density, given by~\eqref{eq:rho_DM}, has non-standard time dependence whenever~$\phi$ evolves. Since the axion evolves at late times from~$\phi \simeq \pi f$ towards~$\phi = 0$, it follows from~\eqref{DM mass axion} 
that~$m(\phi)$ increases at late times. As argued in~\cite{Das:2005yj,Khoury:2025txd} and reviewed below, this gives rise to an effective phantom-crossing behavior.

\subsection{Model parameters and expected values}

Our model is characterized by the following parameters:

\vspace{-0.2cm}
\begin{itemize}

\item The overall scale~$\Lambda$ of the axion potential~\eqref{axion pot full}, which will be~$\sim {\rm meV}$ to achieve late-time acceleration.
It is related to the pion mass~$m_\pi$ and decay constant~$f_\pi$ by
\begin{equation}
\Lambda^4 = \epsilon m_\pi^2 f_\pi^2\,.
\end{equation}
Here,~$\epsilon \ll 1$ is an ad hoc parameter, introduced in~\cite{Hook:2017psm}, that allows the finite-density correction in~\eqref{axion pot full} to compete with the vacuum potential within the perturbative regime.\footnote{To be precise, higher-order density corrections to~\eqref{axion pot full} are~${\cal O}\left( \frac{\sigma_{\rm N}^2n^2}{m_\pi^2f_\pi^2}\right)$, and these are small compared to the leading term as long as
\begin{equation}
\rho_{\rm DM} \lesssim \frac{m_\pi^2 f_\pi^2}{\sigma_{\rm N}/m_{\rm N}} = \frac{\Lambda^4}{\epsilon \sigma_{\rm N}/m_{\rm N}} \,.
\end{equation}
Since~$\sigma_{\rm N}/m_{\rm N} \ll 1$ (as discussed below), this requires~$\epsilon \ll 10^{-12}$ for~\eqref{axion pot full} to be valid up to matter-radiation equality.} The required small~$\epsilon$ values can be achieved naturally with multiple DM copies with~$\mathds{Z}_N$ exchange symmetry~\cite{Hook:2018jle,DiLuzio:2021pxd}. This was investigated recently in~\cite{Delaunay:2026jto}.

\item The dark quark mass ratio~$m_{\rm u}/m_{\rm d}$ (equivalently,~$\xi$).

\item The axion decay constant~$f$, together with~$\xi$, controls the curvature around the maximum/minimum of the potential. We will stick to the sub-Planckian range,~$f \lesssim M_{\rm Pl}$, consistent with quantum gravity expectations~\cite{Banks:2003sx,Arkani-Hamed:2006emk,Rudelius:2015xta}.

\item The sigma term~$\sigma_{\rm N}$ is the quark-mass contribution to the nucleon mass,~$\sigma_{\rm N} \sim \frac{\partial m_{\rm N}}{\partial \ln m_q}$. Its expected parametric dependence in chiral perturbation theory is~$\sigma_{\rm N} \sim \frac{m_\pi^2}{4\pi f_\pi}$, such that the dimensionless coupling parameter~$\sigma_{\rm N}/m_{\rm N}$ satisfies
\begin{equation}
\frac{\sigma_{\rm N}}{m_{\rm N}} \sim \frac{1}{40\pi} \frac{m_\pi^2}{f_\pi^2} \,,
\end{equation}
where we have assumed~$m_{\rm N} = 10 f_\pi$ for concreteness. Thus~$\sigma_{\rm N}/m_{\rm N}$ is naturally small in chiral perturbation theory. In ordinary QCD, for instance, it is given by~$\frac{\sigma_{\rm N}}{m_{\rm N}} = \frac{59~{\rm MeV}}{938~{\rm MeV}}\simeq 0.06$. 

\end{itemize}

\subsection{Background evolution and effective phantom crossing}
\label{background eqns}

Ignoring radiation, the Friedmann equation is\footnote{Here,~$M_{\rm Pl} = \frac{1}{\sqrt{8\pi G_{\rm N}}}$ is the reduced Planck mass.}  
\begin{equation}
    3H^2M_{\rm Pl}^2 = \rho_\phi + \rho_{\rm b} + \rho_{\rm DM}(\phi)\,,
\label{fried}
\end{equation}
where~$\rho_{\rm b}$ is the ordinary baryon density, and~$\rho_\phi  = \frac{1}{2}\dot{\phi}^2 + V(\phi)$ is the usual scalar field energy density.   
Meanwhile, the background scalar equation is
\begin{equation}
\ddot{\phi} + 3H\dot{\phi} = - \frac{{\rm d} V_{\rm eff}}{{\rm d} \phi} \,.
\label{phi eom cosmo}
\end{equation}

To see how these equations lead to effective phantom crossing, suppose that one is fitting data assuming decoupled DM and DE components, 
when the actual underlying model is governed by~\eqref{fried}. In doing so, one is effectively ascribing the non-standard time evolution of DM to the DE density~\cite{Das:2005yj}:
\begin{equation}
    \rho_{\rm DE}^{\rm eff} = \rho_{\phi} +\left [\frac{A(\phi)}{A(\phi_0)}-1\right] \frac{\rho_{\rm DM}^0}{a^3} \,.
\label{rhoDE eff}
\end{equation}
The effective DE density~$\rho_{\rm DE}^{\rm eff}$ is, however, not uniquely specified, since there is an ambiguity in the choice of time at which~$\rho_{\rm DE}^{\rm eff}$ and~$\rho_\phi$ coincide. Equation~\eqref{rhoDE eff} corresponds to coincidence at the present time. 

It is straightforward to show that the effective DE equation of state, defined as~$\frac{{\rm d}\rho_{\rm DE}^{\rm eff}}{{\rm d}t} = - 3H (1 + w_{\rm eff}) \rho_{\rm DE}^{\rm eff} $, is given by
\begin{equation}
    w_{\rm eff} = \frac{w_\phi}{1+\left[\frac{A(\phi)}{A(\phi_0)}-1\right]\frac{\rho_{\rm DM}^0}{a^3\rho_\phi}} \ ,
\label{eq:weff}
\end{equation}
where~$w_\phi = \frac{\dot{\phi}^2 - 2 V(\phi)}{\dot{\phi}^2 + 2V(\phi)}$ is the standard scalar equation of state parameter. Phantom crossing behavior, with~$w_{\rm eff}$ growing from~$< -1$ in the past to~$> -1$ at present, is achieved provided that~$A(\phi) < A(\phi_0)$ at early times. This corresponds physically to a DM mass that increases with time in the redshift range probed by DESI. It is also possible to achieve~$w_{\rm eff} < -1$ with a decreasing DM mass~\cite{Agrawal:2019dlm,Bedroya:2025fwh}, if the coincidence time is set to an earlier redshift instead of the present time.

\subsection{Quintessence and~$\Lambda$CDM limits}\label{sec:limits}

The DADB model has two useful limiting regimes. In the zero-coupling limit,~$\sigma_{\rm N}/m_{\rm N}\rightarrow 0$, the density-dependent correction
to the axion potential vanishes and~$A(\phi)\rightarrow 1$. The model reduces to uncoupled DM and a canonical axion quintessence field.

At the opposite extreme, consider the formal limit~$\sigma_{\rm N}/m_{\rm N}\gg 1$. As the coupling increases, the critical density~\eqref{nc} decreases. Once~$n_{\rm c}<n_0$, the sign flip lies in the future, and the effective potential remains on its high-density branch, whose minimum is at~$\phi=\pi f$, throughout the observed cosmological history. Along this limiting solution, the axion is constant, the DM mass density redshifts as~$a^{-3}$, and~$V(\pi f)$ acts as a cosmological constant. The background evolution is therefore equivalent to that of $\Lambda$CDM.

The perturbations likewise approach the~$\Lambda$CDM limit. At~$\phi=\pi f$, ${\rm d}A/{\rm d}\phi=0$, so the scalar-mediated fifth force vanishes. Moreover, the curvature of the effective potential is
\begin{equation}
m_{\phi,{\rm eff}}^2
\equiv
\left.\frac{{\rm d}^2V_{\rm eff}}{{\rm d}\phi^2}\right|_{\phi=\pi f}
=
\frac{\xi\Lambda^4}{4f^2\sqrt{1-\xi}}
\left(\frac{n}{n_{\rm c}}-1\right) ,
\end{equation}
which is positive on the high-density branch and large when~$n/n_{\rm c}\gg1$. Scalar fluctuations are then heavy and are not sourced
linearly by DM density perturbations, leaving the standard DM perturbation equations. We stress that~$\sigma_{\rm N}/m_{\rm N}\gg 1$ lies outside the expected regime of chiral perturbation theory discussed above. Thus $\Lambda$CDM is a formal limit of the phenomenological DADB model, rather than a controlled limit of its underlying dark-QCD description.

\subsection{Growth of density perturbations}

The DM-DE coupling also impacts the growth of linear density inhomogeneities,
\begin{equation}
\delta_{\rm c} = \frac{\delta\rho_{\rm DM}}{\rho_{\rm DM}}\,.
\end{equation}
This encodes both inhomogeneities in the DM number density, as well as variations in the DM mass.
Working in synchronous gauge, with primes denoting derivatives with respect to conformal time, the equations of motion for linear scalar perturbations are given by~\cite{McDonough:2021pdg,Lin:2022phm}
\begin{eqnarray}
\delta\phi'' +2aH\delta\phi' +\left(k^2+a^2\frac{{\rm d}^2V}{{\rm d}\phi^2}\right)\delta\phi +\frac{1}{2}h'\phi' = \nonumber\\
    -a^2 \left[\frac{{\rm d}\ln A(\phi)}{{\rm d}\phi}\delta_{\rm c} +\frac{{\rm d}^2\ln A(\phi)}{{\rm d}\phi^2}\delta\phi\right]\rho_{\rm DM}(\phi)\,;
\label{pert 1}
\end{eqnarray}
\begin{equation}
    \delta_{\rm c}'+\theta+\frac{h'}{2} = \frac{{\rm d}\ln A(\phi)}{{\rm d}\phi}\delta\phi' +\frac{{\rm d}^2\ln A(\phi)}{{\rm d}\phi^2}\phi'\delta\phi \,;
\label{pert 2}
\end{equation}
\begin{equation}
    \theta'+aH\theta = \frac{{\rm d}\ln A(\phi)}{{\rm d}\phi}k^2\delta\phi -\frac{{\rm d}\ln A(\phi)}{{\rm d}\phi}\phi'\theta \,,
\label{pert 3}
\end{equation}
where~$\theta\equiv\partial_i v^i$ is the DM velocity potential, and~$h$ is the metric trace perturbation in synchronous gauge.
The final equation is that for the metric perturbation arising from the Einstein equations
\begin{equation}
	aH\frac{h'}{2} = k^2\eta + \frac{1}{2M_{\rm Pl}^2}a^2\delta\rho \,;
\label{pert 4}
\end{equation}
\begin{equation}
	\frac{h''}{2} + aH h' - k^2\eta = -\frac{3}{2M^2_{\rm Pl}}a^2\delta P \,,
\label{pert 5}
\end{equation}
where~$\eta$ is the curvature in synchronous gauge, and~$\delta\rho$ and~$\delta P$ are respectively the perturbations to the total density and pressure, including contributions from the dark sector interactions.
We impose adiabatic initial conditions in the deep radiation-dominated era when the scalar field is frozen due to Hubble friction, with~$\delta\phi=\delta\phi'=0$. The CDM velocity is initially set to zero, corresponding to the usual synchronous-gauge choice.

\section{Best-fit model and its interpretation}

Before describing the likelihood analysis of the model (Sec.~\ref{data sec}), it is instructive to
understand physically the behavior of the scalar field evolution, effective equation of state, and growth of
density perturbations for our best-fit model parameters. We focus for concreteness on the CMB+DESI BAO+SNe(DES-Dovekie) combination. 
The best-fit parameters are listed in Table~\ref{tab:param}.

\begin{figure}
    \centering
    \includegraphics[width=0.99\linewidth]{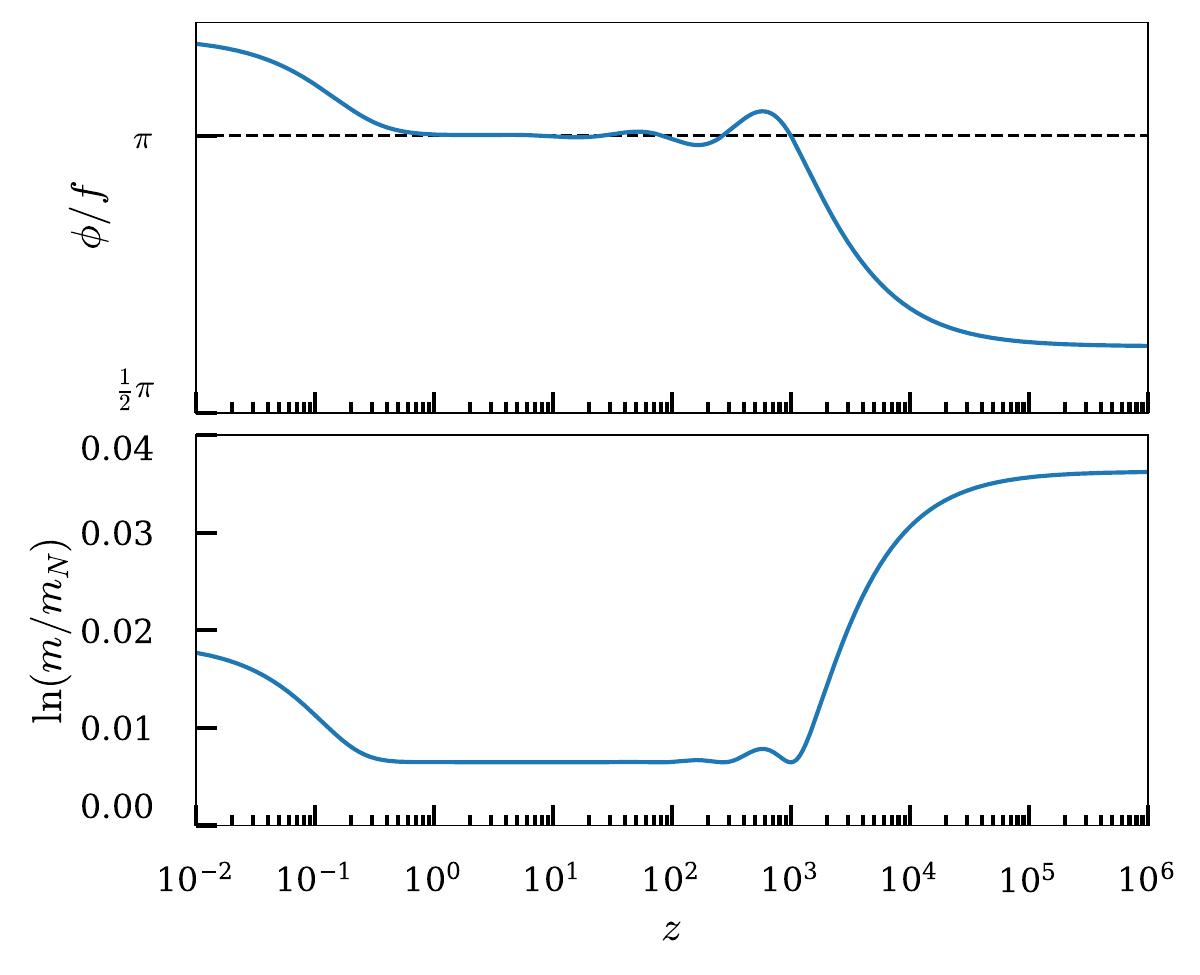}
    \caption{Time evolution of the scalar field (top) and DM mass (bottom) for the best-fit model for the CMB+DESI BAO+SNe(DES-Dovekie) combination.}
    \label{fig:best-phi}
\end{figure}

\begin{figure}
    \centering
    \includegraphics[width=0.99\linewidth]{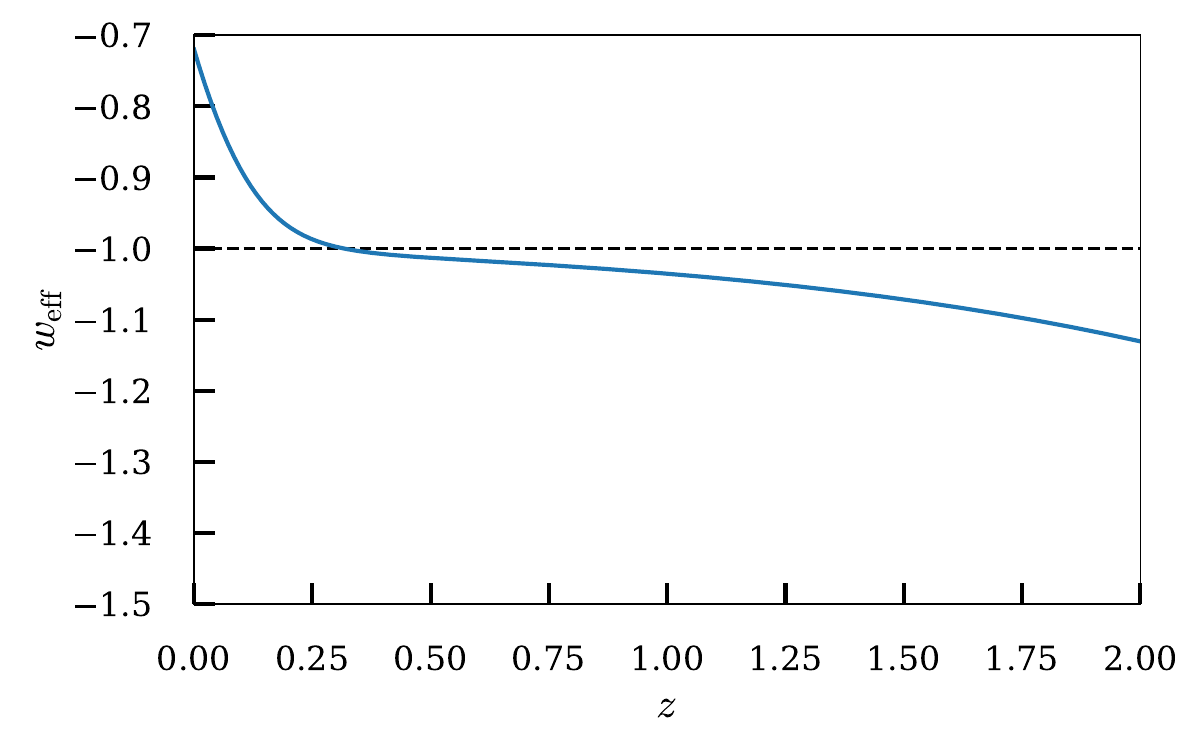}
    \caption{The effective equation of state of DE for the best-fit model for the CMB+DESI DR2+DES-Dovekie data combination. 
    }
    \label{fig:best-weff}
\end{figure}

\subsection{Background evolution}

The background solution for~$\phi$ (Fig.~\ref{fig:best-phi}, top panel) is intuitive. At very early times, the field is frozen at its initial value~$\theta_{\rm i} \equiv \phi_{\rm i}/f$ due to Hubble friction.\footnote{It was shown in~\cite{Burrage:2026pei} that under some circumstances the scalar field can be significantly displaced (``kicked") when it is conformally coupled to a particle species in thermal equilibrium that goes from being relativistic to non-relativistic. However, as noted in~\cite{Burrage:2026pei}, for the case of dark baryons it is not clear that these are ever relativistic and, indeed (as in actual QCD) the dark baryons could well be formed when they are already non-relativistic. In this paper, since we are not focused on the early universe, and do not specify the scale at which the dark QCD transition takes place, we neglect this model-dependent effect.}
It remains frozen there until the Hubble rate drops to a value comparable to the effective scalar mass, {\it i.e.}, when
\begin{equation}
m_{\phi,{\rm eff}}^2(\phi_{\rm i}) \sim H^2\,.
%m_{\phi}^{2}(\phi_{\rm i}) = \left.\frac{{\rm d}^2V_{\rm eff}}{{\rm d}\phi^2}\right\vert_{\phi_{\rm i}} \sim 
\label{release cond}
\end{equation}
Since the DM density is much greater than the critical value~\eqref{nc} at early times,~$n\gg n_{\rm c}$, the density-dependent piece in~\eqref{axion pot full} dominates, hence 
\begin{equation}
m_{\phi,{\rm eff}}^2  \sim 2 \frac{\sigma_{\rm N}}{m_{\rm N}} \frac{\rho_{\rm DM}}{f^2} \,.
\end{equation}
Therefore the field begins to roll when
\begin{equation}
\frac{\sigma_{\rm N}}{m_{\rm N}} \frac{M_{\rm Pl}^2}{f^2} \Omega_{\rm DM} \sim {\cal O}(1)\,,
\end{equation}
with~$\Omega_{\rm DM} = \frac{\rho_{\rm DM}}{3H^2M_{\rm Pl}^2}$. Since~$\sigma_{\rm N}/m_{\rm N} \lesssim 1$ and~$f\lesssim M_{\rm Pl}$, this can naturally occur around the time of matter-radiation equality (when~$\Omega_{\rm DM} \sim 1$)\footnote{It is a generic feature that coupling a scalar field to DM with $A(\phi)\sim 1+g\phi^2$ can naturally trigger the rolling of the field around matter-radiation equality~\cite{Lin:2022phm}.}, as is the case in Fig.~\ref{fig:best-phi}. Indeed, our best-fit values for the CMB+DESI BAO+SNe(DES-Dovekie) combination are~$\sigma_{\rm N}/m_{\rm N} = 0.033$ and~$f = 0.22\, M_{\rm Pl}$, which gives~$\frac{\sigma_{\rm N}}{m_{\rm N}} \frac{M_{\rm Pl}^2}{f^2} \simeq 0.7$. Correspondingly, the DM mass (Fig.~\ref{fig:best-phi}, bottom panel) decreases between matter-radiation equality and recombination, which, as argued in Sec.~\ref{sec:geo}, is necessary to fit the CMB. 

After recombination, the axion rapidly settles to the effective minimum at~$\phi = \pi f$. It remains there until the DM number density reaches the critical value~\eqref{nc}, at which time the sign of the axion potential flips, and~$\phi = \pi f$ becomes a maximum. More precisely, Eq.~\eqref{nc} implies
\begin{equation}
\rho_{\rm DM}^{\rm c} = m_{\rm N} n_{\rm c} = \frac{\Lambda^4}{2\sigma_{\rm N}/m_{\rm N}} \,.
\end{equation}
For~$\Lambda\sim {\rm meV}$ and~$\sigma_{\rm N}/m_{\rm N} \lesssim 1$, the field begins to roll away near the onset of DE domination at~$z \lesssim 1$. Correspondingly, the DM mass increases slightly at late times. The effective equation of state is shown in Fig.~\ref{fig:best-weff} for the best-fit parameter values. Consistent with the discussion in Sec.~\ref{background eqns},~$w_{\rm eff}$ displays phantom crossing at~$z \simeq 0.3$ and evolves to~$w_{\rm eff} \simeq -0.7$ at the present time. Fitting~$w_{\rm eff}$ in the~$z<2$ range with the CPL parameterization gives~$w_0=-0.87$ and~$w_a=-0.36$.\footnote{Because we fit the whole curve with two parameters~($w_0$ and~$w_a$), we find~$w_0\neq w_{\rm eff}(z=0)$.} 

\begin{figure}
    \centering
     \includegraphics[width=0.99\linewidth]{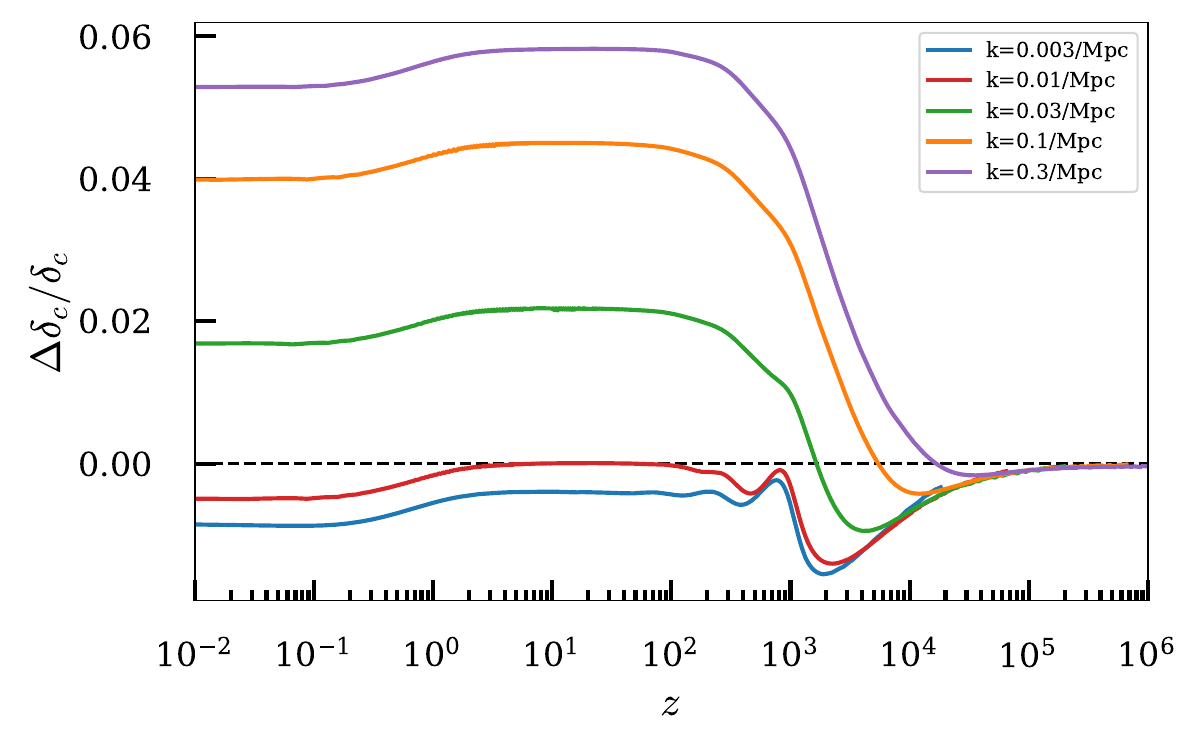}
    \caption{Evolution of the synchronous-gauge DM density perturbation~$\delta_{\rm c}=\delta\rho_{\rm DM}/\rho_{\rm DM}$ relative to the best-fit~$\Lambda$CDM model for the same CMB+DESI DR2+DES-Dovekie data combination. The curves show different comoving wavenumbers, in units of~${\rm Mpc}^{-1}$. The scale dependence reflects whether a mode was inside the horizon during the early epoch of DM mass decrease.}
    \label{fig:best-deltac}
\end{figure}

\begin{figure}
    \centering
    \includegraphics[width=0.99\linewidth]{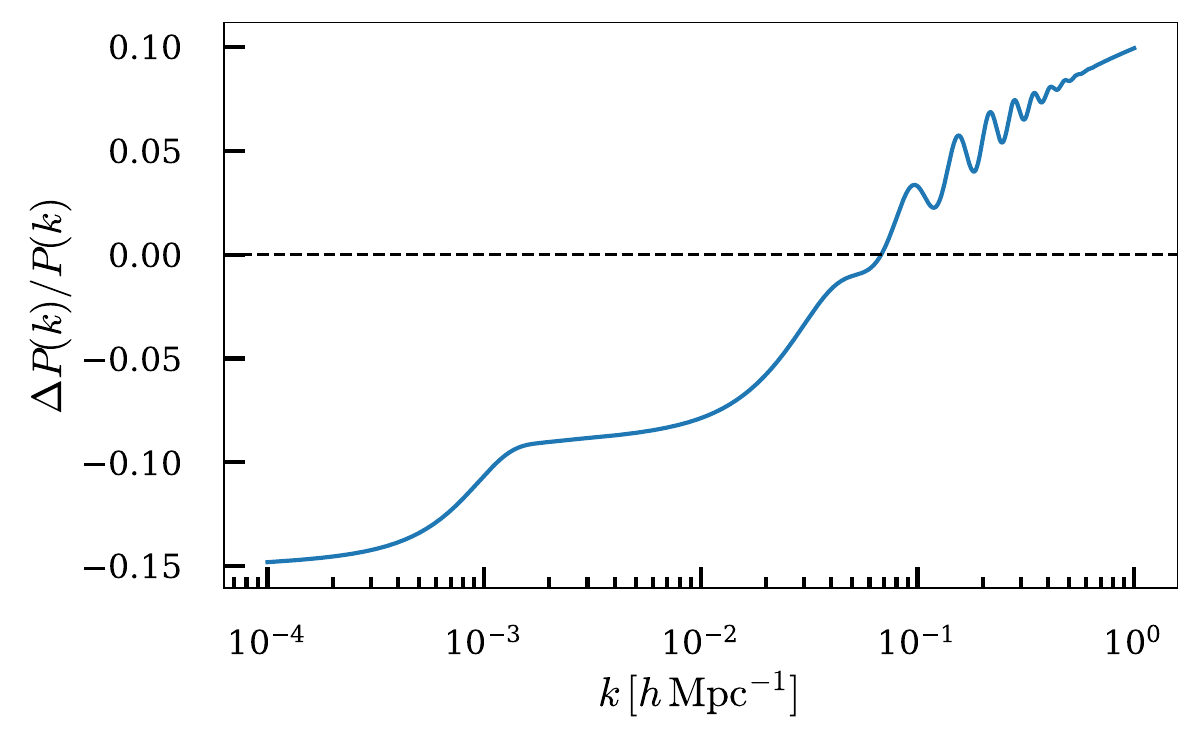}
    \caption{Linear matter power spectrum at~$z=0$ for the best-fit DADB model relative to the best-fit~$\Lambda$CDM model for the same default data combination. The large-scale suppression and small-scale enhancement are the transfer-function imprint of the scale-dependent growth shown in Fig.~\ref{fig:best-deltac}.}
    \label{fig:best-Pk}
\end{figure}

\subsection{Density perturbations}

Let us next discuss the evolution of the synchronous-gauge DM density perturbation~$\delta_{\rm c}=\delta\rho_{\rm DM}/\rho_{\rm DM}$ for our
best-fit model, obtained by numerically integrating Eqs.~\eqref{pert 1}-\eqref{pert 5}. Figure~\ref{fig:best-deltac} shows
the fractional change relative to a fiducial~$\Lambda$CDM model,
\begin{equation}
\frac{\Delta \delta_{\rm c}}{\delta_{\rm c}}
=
\frac{\delta_{\rm c}-\delta_{\rm c}^{\Lambda{\rm CDM}}}
{\delta_{\rm c}^{\Lambda{\rm CDM}}}\,,
\end{equation}
on several scales as a function of redshift. The fiducial model is the best-fit~$\Lambda$CDM cosmology for the same data combination.

The scale dependence can be understood directly from the Euler equation (Eq.~\eqref{pert 3}), which can be written as
\begin{equation}
    \theta'+\Big[aH+(\ln A)'\Big]\theta = \frac{{\rm d}\ln A}{{\rm d}\phi} k^2\delta\phi\,.
    \label{eq:euler_mass_friction}
\end{equation}
Thus the time evolution of the DM mass modifies the effective Hubble drag on peculiar velocities. During the early epoch in which the DM mass decreases, $(\ln A)'<0$, and the damping term in Eq.~\eqref{eq:euler_mass_friction} is reduced. Modes that are already inside the horizon during this period have appreciable velocity divergence and therefore receive a growth-rate boost. Superhorizon modes are much less sensitive to this effect, because their velocity is gradient-suppressed.\footnote{For regular adiabatic initial conditions,~$\theta$ starts at order~$k^2$ in the gradient expansion. This is also manifest in Eq.~\eqref{eq:euler_mass_friction}: the scalar-force source is proportional to~$k^2\delta\phi$, while the friction term only damps or amplifies an already existing~$\theta$. Thus, superhorizon modes are much less affected by the temporary reduction of the effective Hubble drag.}

The scalar force term on the right-hand side of Eq.~\eqref{eq:euler_mass_friction} provides an additional scale-dependent effect. It is most efficient for modes inside the scalar Compton wavelength,~$k/a\gtrsim m_{\phi,{\rm eff}}$, and is suppressed on larger scales. In addition, the best-fit DADB solution has a slightly larger physical DM density around matter-radiation equality than the fiducial~$\Lambda$CDM model, shifting equality earlier and reducing the usual radiation-era suppression of modes that entered the horizon before equality. These effects explain why small-scale modes are enhanced relative to~$\Lambda$CDM, whereas larger-scale modes do not receive the early boost.

At late times, the modified DE expansion history and the subsequent increase of~$A$ mildly suppress growth on all scales. The largest-scale modes, which did not benefit from the early subhorizon enhancement, are therefore suppressed in the final matter power spectrum. We show this spectrum relative to the best-fit~$\Lambda$CDM model in Fig.~\ref{fig:best-Pk}. The present analysis does not include late-time galaxy clustering, weak-lensing, or redshift-space distortion measurements. A full comparison with these data, including the treatment of nonlinear scales, is left for future work; see Ref.~\cite{Costa:2025kwt} for a related discussion.

\begin{figure}
    \centering
    \includegraphics[width=0.99\linewidth]{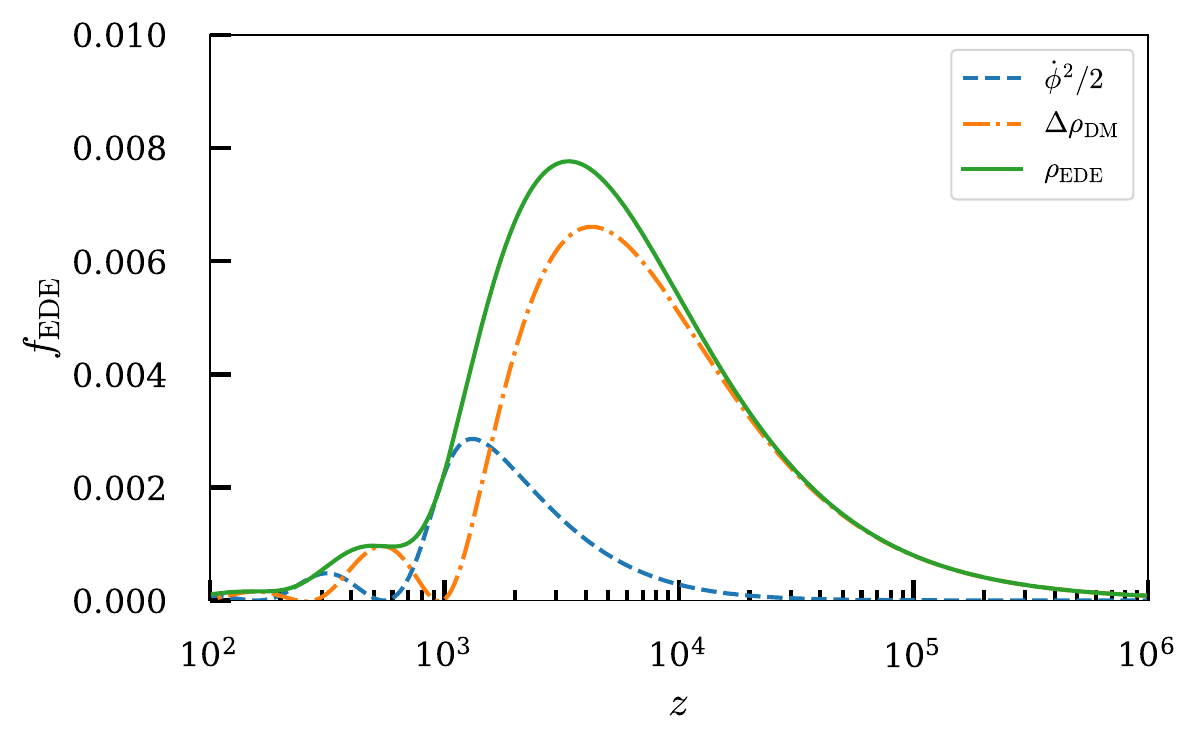}
    \caption{Fractional energy density of the Early Dark Energy component for the best-fit model for the CMB+DESI DR2+DES-Dovekie data combination. The blue dashed and orange dash-dotted lines denote the contributions from the scalar kinetic energy and the excess DM mass density, respectively, see Eq.~\eqref{eq:rhoEDE}. The green solid line shows their sum,~$f_{\rm EDE}(z)$.  
    }
    \label{fig:fEDE}
\end{figure}

\subsection{Unified early- and late-time dark energy}
\label{sec:EDE}

The DADB model was originally motivated as a well-behaved field-theoretic origin of apparent phantom crossing. Interestingly, the same dynamics also
generates a transient early dark energy (EDE)-like contribution, thereby providing a framework in which early and late-time DE arise from a single
interacting dark sector. Similar behavior has been noticed in other interacting DM-DE models, {\it e.g.},~\cite{Smith:2025grk,Andriot:2025los,Giare:2026tyk}.

Models that substantially alleviate the Hubble tension typically require a transient contribution of order~$\sim 10\%$ to the total energy density near matter-radiation equality, followed by rapid dilution before late times~\cite{Poulin:2018cxd,Lin:2019qug}. Unifying such a component with late-time DE is challenging because of the large hierarchy between the two relevant energy scales. In the DADB model, this hierarchy is bridged by the cosmological evolution of the DM density: late-time acceleration is controlled by the vacuum axion potential, whereas the early-time contribution is set by the density-dependent axion--dark-baryon interaction. Whether the resulting contribution is large enough to significantly affect the Hubble tension is a quantitative question that we address below.

To isolate the early-time contribution, we take as a reference the constant DM mass attained at the high-density minimum,~$m_{\rm N}A(\pi f)$. During the pre-recombination epoch, when the vacuum axion potential is negligible, we define
\begin{equation}
 \rho_{\rm EDE}
 \equiv
 \frac{1}{2}\dot{\phi}^{\,2}
 +m_{\rm N} \big[A(\phi)-A(\pi f)\big] n\,.
 \label{eq:rhoEDE}
\end{equation}
The two terms represent, respectively, the scalar kinetic energy and the excess DM mass density relative to its value at~$\phi=\pi f$. Although the separation of an interacting dark sector into individual components is not unique, this definition isolates the transient equality-era contribution and gives~$\rho_{\rm EDE}\rightarrow0$ once the field settles at the high-density minimum.
%Including the vacuum contribution would add~$V(\phi)-V(\pi f)$ to Eq.~\eqref{eq:rhoEDE}, which is negligible during the epoch of interest.

Near~$\phi=\pi f$, these contributions together describe the energy stored in the oscillating scalar mode. 
For~$n\gg n_{\rm c}$, its effective mass satisfies~$m_{\phi,{\rm eff}}^2\propto n\propto a^{-3}$,
and therefore~$m_{\phi,{\rm eff}}\propto a^{-3/2}$. In the adiabatic oscillatory regime, the comoving number of scalar quanta is conserved,
so the corresponding physical number density satisfies~$n_\phi^{\rm osc}\propto a^{-3}$. Therefore,
\begin{equation}
  \rho_{\rm EDE}\simeq m_{\phi,{\rm eff}} \,n_\phi^{\rm osc}
  \propto a^{-3/2}a^{-3}=a^{-9/2}\,.
\end{equation}
Hence this early-time contribution dilutes faster than radiation, thereby realizing the desired behavior of an EDE component.
Whether the DADB model can simultaneously resolve the Hubble tension and the apparent phantom-crossing behavior depends on the overlap of the parameter regions required to address each tension.

We characterize this episode by the redshift-dependent fractional contribution
\begin{equation}
 f_{\rm EDE}(z)
 \equiv
 \frac{\rho_{\rm EDE}(z)}{\rho_{\rm tot}(z)}
 =
 \frac{\rho_{\rm EDE}(z)}{3M_{\rm Pl}^{2}H^{2}(z)}\,.
 \label{eq:fEDE}
\end{equation}
We denote its early-time peak by~$f_{\rm EDE}^{\rm peak}\equiv f_{\rm EDE}(z_{\rm EDE})$, where~$z_{\rm EDE}$ is the redshift at which this local maximum occurs. For the best-fit model, we find~$f_{\rm EDE}^{\rm peak}=0.0076$ at~$z_{\rm EDE}\simeq 3.5\times10^3$, as shown in Fig.~\ref{fig:fEDE}. At current confidence levels, this falls short of fully addressing the Hubble tension, which requires~$f_{\rm EDE} \sim 0.1$. 

\section{Data Analysis}
\label{data sec}

\begin{figure*}
    \centering
    \includegraphics[width=0.99\linewidth]{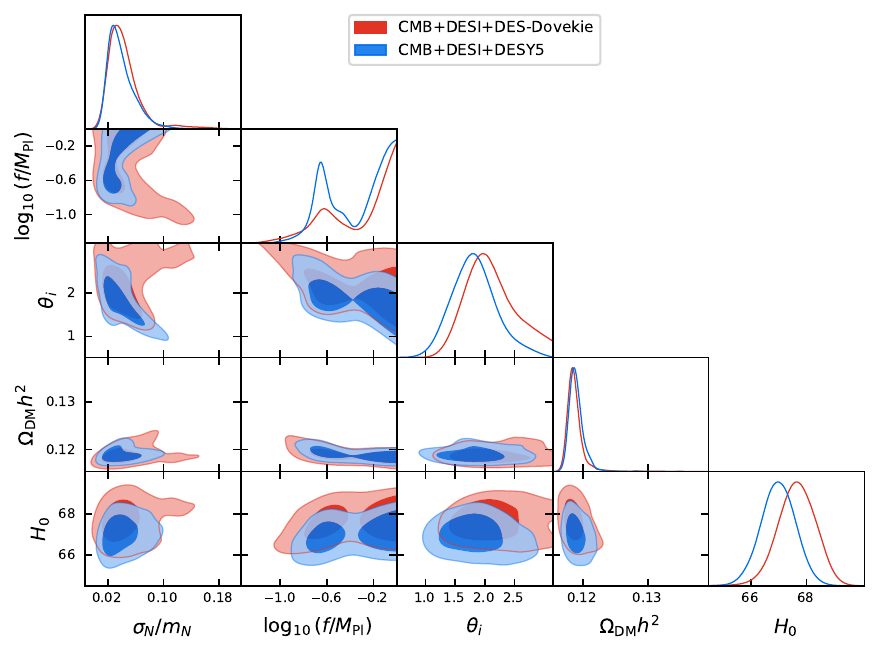}
    \caption{
    The marginalized joint posterior of parameters of the DADB model for different datasets.
    The darker and lighter shades correspond respectively to the 68\% C.L. and 95\% C.L.
    }
    \label{fig:posterior}
\end{figure*}

\begin{table*}[]
    \centering
    \begin{tabular}{c||c|c|c|c|c}
        \hline
        Data & CMB+DESI & +DES-Dovekie & +DESY5 & +Union3.1 & +Corr. Pantheon+ \\
        \hline
        $\sigma_{\rm N}/m_{\rm N}$ & 0.032 & 0.033 ($0.044^{+0.010}_{-0.026}$) & 0.032 ($0.039^{+0.010}_{-0.023}$) & 0.033 & 0.032 \\
        $f/M_{\rm Pl}$             & 0.22  & 0.22 ($>0.30$)                    & 0.22 ($>0.28$)                    & 0.22 & 0.22 \\
        $\theta_{\rm i}=\phi_{\rm i}/f$        & 1.91  & 1.95 ($2.10^{+0.32}_{-0.51}$)     & 1.92 ($1.82^{+0.32}_{-0.41}$)     & 1.88 & 1.95 \\
        \hline
        $H_0$ (${\rm km/s/Mpc}$)   & 66.45 & 67.42 ($67.65\pm0.68$)            & 66.88 ($66.98\pm0.63$)            & 66.89 & 67.62 \\
        $\Omega_{\rm DM}h^2$ & 0.1204 & 0.1199 ($0.1189^{+0.0004}_{-0.0014}$) & 0.1203 ($0.1190^{+0.0007}_{-0.0012}$) & 0.1203 & 0.1198 \\
        $\Omega_{\rm b}h^2$            & 0.0226 & 0.0226 ($0.0224\pm 0.0002$)       & 0.0226 ($0.0224\pm 0.0002$)       & 0.0226 & 0.0225 \\
        $\ln(10^{10}A_s)$          & 3.05  & 3.05 ($3.05\pm0.01$)              & 3.05 ($3.05\pm0.01$)              & 3.05 & 3.05 \\
        $n_s$                      & 0.973 & 0.973 ($0.968\pm 0.004$)          & 0.973 ($0.968\pm 0.004$)          & 0.973 & 0.973 \\ 
        $\tau_{\rm reio}$          & 0.054 & 0.053 ($0.057\pm 0.008$)          & 0.053 ($0.056\pm 0.008$)          & 0.053 & 0.053 \\
        \hline
    \end{tabular}
    \caption{Best-fit DADB parameters for different SNe samples. For the two data combinations with full MCMC chains, marginalized 68\% credible intervals are shown in parentheses.
    }
    \label{tab:param}
\end{table*}

\begin{table*}[]
    \centering
    \begin{tabular}{l|c|c|c|c|c}
        \hline
        \multicolumn{1}{l|}{Model: data} & $\Delta N_{\rm param}$ & $\Delta\chi^2_{\rm total}$ & $\Delta\chi^2_{\rm CMB}$ & $\Delta\chi^2_{\rm DESI}$ & $\Delta\chi^2_{\rm SNe}$   \\
        \hline
        \hline
        DADB: CMB+DESI+DES-Dovekie          & 3 & -14.48 & -5.12 & -2.77 & -6.53   \\
        DADB: CMB+DESI+Union3.1            & 3 & -14.89 & -7.64 & -2.02 & -5.15   \\
        DADB: CMB+DESI+Corr. Pantheon+      & 3 & -15.83 & -9.12 & -1.42 & -5.15 \\
        DADB: CMB+DESI                     & 3 & -8.94  & -5.77 & -2.27 & --       \\
        \hline
        $w_0w_a$: CMB+DESI+DES-Dovekie      & 2 & -12.43 & -4.32 & -3.16 & -5.51 \\
        $w_0w_a$: CMB+DESI+Union3.1        & 2 & -13.17 & -7.09 & -2.10 & -4.25   \\
        $w_0w_a$: CMB+DESI+Corr.Pantheon+  & 2 & -13.87 & -7.19 & -1.97 & -5.44   \\
        $w_0w_a$: CMB+DESI                 & 2 & -8.44 & -4.27 & -3.75 & --   \\
        \hline
    \end{tabular}
    \caption{Comparison between the maximum likelihood models to different data combinations. The~$\Delta N_{\rm param}$ is the number of extra parameters in addition to the~$\Lambda$CDM ones. The~$\Delta\chi^2$ are calculated compared to the $\Lambda$CDM model fitted to the same datasets, and $\Delta\chi^2_{\rm total}$ includes the contributions from the prior of nuisance parameters which are not shown here.
    }
    \label{tab:chi2}
\end{table*}

\begin{table*}
    \centering
    \begin{tabular}{l|c|c|c|c|c}
        \hline
        \multicolumn{1}{l|}{Model: data} & $\Delta N_{\rm param}$ & $\Delta\chi^2_{\rm total}$ & $\Delta\chi^2_{\rm CMB}$ & $\Delta\chi^2_{\rm DESI}$ & $\Delta\chi^2_{\rm SNe}$   \\
        \hline
        \hline
        DADB: CMB+DESI+DESY5               & 3 & -22.21 & -4.91 & -3.92 & -12.87 \\
        DADB: CMB+DESI+Union3              & 3 & -17.32 & -5.06 & -2.59 & -7.67  \\
        DADB: CMB+DESI+Pantheon+           & 3 & -10.18 & -3.55 & -2.87 & -2.75  \\
        \hline
        $w_0w_a$: CMB+DESI+DESY5           & 2 & -19.23 & -4.02 & -4.04 & -11.23 \\
        $w_0w_a$: CMB+DESI+Union3          & 2 & -15.18 & -4.70 & -2.93 & -6.57  \\
        $w_0w_a$: CMB+DESI+Pantheon+       & 2 & -8.89  & -2.67 & -3.30 & -2.54  \\
        \hline
    \end{tabular}
    \caption{Similar to Table~\ref{tab:chi2} but with original-calibrated SNe samples.
    }
    \label{tab:chi2-original}
\end{table*}

\subsection{Method}

We implement the DADB model with consistent linear perturbations using a modified version of CLASS~\cite{2011arXiv1104.2932L,2011JCAP...07..034B} and perform Markov chain Monte Carlo (MCMC) analyses using Cobaya~\cite{Torrado:2020dgo}. Following the procedure used for the standard flat~$\Lambda$CDM model, we fix the curvature~$\Omega_K=0$, the effective number of relativistic species~$N_{\rm eff}=3.046$, and the sum of the neutrino masses~$\sum m_\nu=0.06~{\rm eV}$. For the dark sector parameters, the quark mass ratio is fixed to~$m_{\rm u}/m_{\rm d}=0.8$, while the DE scale~$\Lambda$ is set by requiring the Friedmann equation~\eqref{fried} to be satisfied today, given the input values of~$H_0$, and the baryon and DM densities.

We sample the bare dark-baryon density
\begin{equation}
\omega_{\rm N}\equiv \frac{m_{\rm N}n_0}{3M_{\rm Pl}^2H_{100}^2}\,, 
\end{equation}
with~$H_{100}\equiv100\,{\rm km\,s^{-1}\,Mpc^{-1}}$. The physical present-day CDM density is then the derived quantity~$\omega_c=A(\phi_0)\omega_{\rm N}$, where~$\phi_0$ is obtained by solving the background equations. This avoids the need to iteratively determine the initial number density for each proposed set of parameter values.

Thus, in addition to the standard six~$\Lambda$CDM parameters, we have three more parameters with flat priors: $\sigma_{\rm N}/m_{\rm N}\in [0,0.5]$, $\log_{10}(f/M_{\rm Pl})\in [-3,0]$, and $\theta_{\rm i}\in [0,3.14159]$. The arbitrary upper limit of $\sigma_{\rm N}/m_{\rm N}$ is only valid when a non-zero coupling is significantly preferred over the $\Lambda$CDM limit, as we show later. We impose a sub-Planckian prior on~$f$, motivated by quantum-gravity considerations for axion decay constants~\cite{Banks:2003sx,Arkani-Hamed:2006emk}. Since the posterior has support near this boundary, the dependence of the constraints on the upper prior for~$f$ should be kept in mind.

We define the labels and details of the datasets used in this paper as:
\begin{itemize}
    \item CMB: Planck 2018 high-$\ell$ TTTEEE, low-$\ell$ TT, EE, and CMB lensing likelihoods~\cite{Planck:2018vyg}.
    \item DESI: DESI DR2 BAO measurements~\cite{DESI:2025zgx} including the bright galaxy sample (BGS), luminous red galaxies (LRGs), emission line galaxies (ELGs), and quasars (QSOs) with their Lyman-$\alpha$ forests.
    \item DESY5: DES Year-5 SNe Ia sample with original calibration~\cite{DES:2024jxu}.
    \item DES-Dovekie: DES-Dovekie recalibrated year-5 SNe Ia samples~\cite{DES:2025sig}.
    \item Pantheon+: Pantheon+ SNe Ia sample~\cite{Scolnic:2021amr}.
    \item Corr. Pantheon+: Pantheon+ SNe Ia sample after correction of the host galaxy property dependence~\cite{Hoyt:2026fve}.
    \item Union3: Union3 SNe Ia sample~\cite{Rubin:2023ovl}.
    \item Union3.1: Union3.1+UNITY1.8 SNe Ia sample after correction of the host galaxy property dependence~\citep{Rubin:2026qdt,Hoyt:2026fve}.
    %\item H0: Prior on $H_0$ from SH0ES 2020 measurement $H_0=73.2\pm1.3\,{\rm km/s/Mpc}$~\cite{Riess:2020fzl}.
\end{itemize}
Our default combination is CMB+DESI+DES-Dovekie. We also test the combinations with the other SNe data.

%%%%%%%%%%%
\subsection{Results}\label{sec:results}
We show the posterior distributions of the DADB model fitted to our default dataset combination in Fig.~\ref{fig:posterior}, together with the corresponding results obtained using DESY5 instead of DES-Dovekie for comparison. The posterior distributions are qualitatively similar for the two data set combinations.  
The marginalized 95\% credible region for the coupling parameter, $\sigma_{\rm N}/m_{\rm N}$, excludes zero at the 95\% confidence level and exhibits a non-Gaussian tail extending toward larger values. The axion decay constant reaches the physical upper bound, $f\leq M_{\rm Pl}$, and displays a bimodal posterior distribution. The best-fit parameters and marginalized 68\% confidence intervals are summarized in Table~\ref{tab:param}.

Because the posterior distributions are significantly non-Gaussian, it is not straightforward to quantify either the preference for a nonzero coupling or the deviation from the $\Lambda$CDM limit (corresponding to the infinite-coupling limit; see Sec.~\ref{sec:limits}) using marginalized constraints alone. Instead, Table~\ref{tab:chi2} reports the improvement in the best-fit~$\chi^2$ relative to $\Lambda$CDM for the same dataset combinations. For our default dataset, the DADB model improves the fit over $\Lambda$CDM by~$\Delta\chi^2_{\rm total}=-14.48$ with three additional parameters, yielding better fits to the CMB, BAO, and SNe datasets individually. 

Similar improvements are obtained when replacing the SNe sample with other recalibrated or corrected datasets, such as Union3.1 and Corr. Pantheon+~\cite{Hoyt:2026fve}. Even without including SNe data, the DADB model still improves the fit by~$\Delta\chi^2_{\rm total}=-8.94$ relative to $\Lambda$CDM. Compared with the phenomenological~$w_0w_a$CDM model, the DADB model provides an additional improvement of approximately~$\Delta\chi^2_{\rm total}\simeq-2$, despite introducing only one extra parameter, while offering a well-motivated particle-physics interpretation. For completeness, we also present the best-fit results obtained using the original calibrated SNe datasets (DESY5, Union3, and Pantheon+) in Table~\ref{tab:chi2-original}.

Although the DADB model has the potential to alleviate the Hubble tension through the emergence of an EDE component, as discussed in Sec.~\ref{sec:EDE}, the EDE fraction preferred by current data is too small to fully resolve the tension (Fig.~\ref{fig:fEDE}). We find~$H_0=67.65\pm 0.68\,{\rm km/s/Mpc}$ for the CMB+DESI+DES-Dovekie dataset, and~$H_0=66.98\pm 0.63\,{\rm km/s/Mpc}$ when DESY5 is used instead of DES-Dovekie. The former yields a slightly higher value of~$H_0$, but nevertheless remains in~$4.4\sigma$ tension with the SH0ES local distance-ladder measurement~\cite{Riess:2021jrx}.

%%%%%%%%%

\section{Conclusions} \label{sec:conclusion}
We have investigated the viability of the dark axion--dark baryon (DADB) interaction model proposed in~\cite{Khoury:2025txd} by confronting its predictions with the latest cosmological observations. For our default dataset combination CMB+DESI DR2 BAO+SNe(DES-Dovekie), we find evidence for nonzero dark-sector interactions, with an improvement of $\Delta\chi^2=-14.48$ over $\Lambda$CDM despite the introduction of only three additional parameters. Unlike phenomenological extensions, the DADB model also provides a well-motivated particle-physics framework for interpreting the data. Comparisons with alternative dataset combinations and with the phenomenological $w_0w_a$CDM model are summarized in Table~\ref{tab:chi2}.

In the DADB model, the DM mass decreases between matter-radiation equality and recombination, and subsequently increases during the BAO- and SNe-sensitive epoch. This evolution allows the model to remain consistent with CMB observations while naturally producing an apparent DE phantom-crossing behavior at late times. Interestingly, an EDE component emerges around matter-radiation equality, opening the possibility of a unified description of early- and late-time DE within a single framework. Although this mechanism can raise the inferred value of $H_0$, the EDE fraction preferred by current data is not sufficiently large to fully resolve the Hubble tension.

The best-fit DADB model also predicts slightly enhanced matter growth compared with the best-fit $\Lambda$CDM model. However, since late-time galaxy clustering, weak lensing, and redshift-space distortion measurements are not included in the present analysis, a comprehensive assessment of structure growth in this model is left for future work.

Future cosmological observations, particularly improved BAO, SNe, and large-scale structure measurements, will provide stringent tests of this scenario. 
The scalar-mediated DM self-interactions predicted by the model may provide an independent cross-check, complementary to the cosmological tests studied here.

\textbf{Acknowledgments:} 
We thank Lam Hui for helpful discussions and Taylor Hoyt and David Rubin for sharing the recalibrated SNe data.
The work of J.K. and M.T. is supported in part by the DOE (HEP) Award No. DE-SC0013528. M-X.L. is supported by funds partially provided by the Canadian Institute for Theoretical Astrophysics (CITA) National Fellowship and funds provided by the Center for Particle Cosmology.
Computing resources were provided by the University of Chicago Research Computing Center through the Kavli Institute for Cosmological Physics at the University of Chicago.

\bibliography{ref}

%\clearpage

%\appendix

\end{document}